\documentclass[a4paper,11pt]{article}
\pdfoutput=1
\usepackage{jheppub}
\usepackage{amsmath,amssymb}

\newcommand{\be}{\begin{equation}}
\newcommand{\ee}{\end{equation}}
\newcommand{\rep}[1]{\mathbf{#1}}
\newcommand{\brep}[1]{\overline{\mathbf{#1}}}

\title{\boldmath The complete massless singlet spectrum in the free-field construction of heterotic strings on Calabi--Yau orbifolds}

\author[a]{Vladimir Belavin}
\affiliation[a]{Physics Department, Ariel University, Ariel 40700, Israel}
\emailAdd{vladimirbe@ariel.ac.il}

\abstract{We apply the free-field construction of the heterotic string compactified on Berglund--H\"ubsch Calabi--Yau orbifolds by computing the full spectrum of massless $E_6$ singlets. The contribution of the descendant vertices is obtained by combining the exact content of the irreducible $N=2$ minimal model representations, encoded in the ranks of the Shapovalov matrices, with the twisted sector structure of the orbifold. The method reproduces the known spectrum of the quintic orbifold with Hodge numbers $(17,21)$, namely $17$ generations, $21$ antigenerations and $234$ singlets. For the quintic itself we obtain $330$ singlets. We show that this number, rather than the frequently quoted value $326$, obtained from the geometric description, is the correct one at the Gepner point, in agreement with the Landau--Ginzburg computation of Kachru and Witten. We also provide an explicit construction of the general vertices. We then compute the new spectra of the two remaining quintic orbifolds, $Z_5[0,1,2,3,4]$ with $(21,1,210)$ and $Z_5[0,0,0,1,4]$ with $(49,5,258)$, where the exceptional Hodge number $h^{2,1}=49$ arises from the twisted sectors. All four examples satisfy exact mirror-symmetry checks.}

\begin{document}
\maketitle
\flushbottom

\section{Introduction}
\label{sec:intro}

Compactification of the ten-dimensional heterotic string on a Calabi--Yau (CY) threefold provides the standard framework for constructing four-dimensional models with $N=1$ spacetime supersymmetry and $E_6$ gauge symmetry \cite{Candelas:1985en}. The massless matter consists of generations in the $\rep{27}$ and antigenerations in the $\brep{27}$ representation of $E_6$, counted by the Hodge numbers $h^{2,1}$ and $h^{1,1}$ respectively, and of $E_6$ singlets. The singlets contain the moduli of the compactification together with the deformations of the tangent bundle, and their number is not a topological invariant. An exact description of all three sectors is available when the internal $N=2$ superconformal theory with $c=9$ is solvable. The classical example is the Gepner construction \cite{Gepner:1987qi,Gepner:1988qi}, which represents the internal theory as an orbifold of a product of $N=2$ minimal models and is applicable when the CY manifold is of Fermat type. The corresponding point of the moduli space, where the internal theory is this exactly solvable conformal field theory, is called the Gepner point. It is a special, symmetric point, distinct from a generic point of the moduli space, where the compactification is described by classical geometry and the massless singlets are counted geometrically.

A free-field construction of the heterotic string on the much larger class of Berglund--H\"ubsch (BH) Calabi--Yau manifolds \cite{Berglund:1993fj} was proposed in \cite{Belavin:2025ff1} and developed for orbifolds in \cite{Makarov:2026ff2}, using the Batyrev--Borisov combinatorial approach \cite{Batyrev:1994hm,Borisov:2013}. In this construction the physical states are vertex operators built from free bosons and fermions. Their momenta are quantized on a pair of dual lattices $M(G)$ and $N(G)$ determined by the orbifold group $G$, the physical subspace is selected by the Borisov cohomology together with the GSO conditions, and complete vertices are products of left and right factors subject to mutual locality. The construction reproduces the counting of $\rep{27}$ and $\brep{27}$ representations by the points of the Batyrev polytopes, and for Fermat models it is connected to the Gepner construction by an explicit dictionary \cite{Makarov:2026ff2,Belavin:2023bp}.

The singlet sector is more subtle. Recall that in the heterotic string the two chiralities are distinct: the right-moving sector is bosonic and carries the gauge quantum numbers, while the left-moving sector is supersymmetric. Here and below $\Delta$ denotes the conformal dimension and $Q$ the charge of the $U(1)$ R-current of the internal $N=2$ algebra. On the right-moving side this $U(1)$ is the abelian factor in $E_6\supset SO(10)\times U(1)$, so the charge $Q$ of a right factor enters the $E_6$ quantum numbers of the corresponding state, and an $E_6$ singlet must be neutral under it. Accordingly, the generations and antigenerations are built on charged right factors, with $(\Delta,Q)=(\tfrac12,+1)$ for the $\rep{27}$ and $(\tfrac12,-1)$ for the $\brep{27}$. A complete singlet is a product of a left vertex with $(\Delta,Q)=(\tfrac12,+1)$ and a right vertex with $(\Delta,Q)=(1,0)$ in the CY sector. The left factor is rigid, it is a chiral primary. On the right side, however, the exponential vertices alone do not exhaust the states, and the spectrum contains the so-called general vertices, descendants dressed by oscillator prefactors. 

Examples of these states were considered in \cite{Belavin:2025ff1} and \cite{Makarov:2026ff2}, but no closed rule was introduced. The difficulty has a simple origin. An exponential vertex specifies a position inside an irreducible representation, i.e., the level and the charge, but not the number of independent states at that position. This number can be zero when a null vector occurs, one, or larger, and it cannot be read from the exponential alone. For this reason the total numbers of singlets quoted in \cite{Belavin:2025ff1} and \cite{Makarov:2026ff2} cannot be reproduced without an additional counting rule, and this blocks the computation of new examples, in particular of the so-called exceptional orbifolds. For a generic BH orbifold the Hodge numbers equal the numbers of invariant deformations of the potential and of the mirror potential. In the exceptional cases the true Hodge numbers are larger, the excess originating in the twisted sectors, and the deformation counting fails. The refined combinatorial formula for the exceptional Hodge numbers, in terms of the so-called Roan pairs of group elements, has recently appeared in \cite{Aleshin:2026abk}. The present work is complementary to that result. It provides the physical realization of the same states within the free-field construction and extends the counting from the Hodge numbers to the complete massless spectrum, including the singlet sector, which is not addressed by the combinatorial methods.

Our result can be stated as a simple universal rule: the massless spectrum is counted by the ranks of the Shapovalov matrices of the $N=2$ minimal models, evaluated at the grades selected by the free-field data. The counting becomes elementary once it is organized representation by representation. Each factor of the internal theory is an $N=2$ minimal model, so every right vertex belongs to a definite irreducible representation of the $N=2$ algebra and occupies a definite position inside it, labeled by the level $n$ above the primary state and by the shift $\delta$ of the $U(1)$ charge. The number of states at the grade $(n,\delta)$ equals the rank of the Shapovalov matrix of the Verma module at that grade. For the massless problem only levels up to one contribute, the matrices are at most three by three, and unitarity guarantees that all null vectors, including the model-dependent embedded ones, are subtracted automatically. The twisted sectors and the projection are taken in the form given in \cite{Belavin:2022sf,Belavin:2023bp}, and the free-field dictionary translates every state back to an explicit vertex operator.

The plan of the paper is the following. In section \ref{sec:construction}, we recall the free-field construction and formulate the counting problem. In section \ref{sec:modules}, we compute the descendant content of the minimal-model representations from the Shapovalov form. We assemble the complete massless spectrum from the twisted sectors and the locality conditions in section \ref{sec:sectors}. We reproduce the spectrum of the $(17,21)$ orbifold of the quintic, including all $234$ singlets in section \ref{sec:1721}.  In section \ref{sec:quintic}, we revisit the quintic  and obtain $330$ singlets. This number refers to the Gepner point, where the gauge symmetry is enhanced by an extra $U(1)^4$, while the frequently quoted $326$ is the geometric count at generic complex structure, and the two differ by four states charged under the extra $U(1)$ factors. We compute the new spectra of the two remaining quintic orbifolds and verify mirror symmetry in section \ref{sec:new}.

\section{The free-field construction and the counting problem}
\label{sec:construction}

A BH manifold is the zero locus in a weighted projective space of a quasi-homogeneous polynomial $W_0$ built of five monomials, so that $W_0(\lambda^{\rho_i}x_i)=\lambda^d W_0(x_i)$, where $\rho_i$ are the weights and the CY condition reads $d=\sum_i \rho_i$. For the quintic $W_0=\sum_i x_i^5$, all $\rho_i=1$ and $d=5$. An orbifold is specified by an admissible group $G$ of phase symmetries $x_i\to e^{2\pi i w_i/d}x_i$ preserving the holomorphic three-form, and the BH mirror manifold is governed by the dual group $G^*$ \cite{Berglund:1993fj,Krawitz:2009}.

In the heterotic string the internal theory enters twice. The right-moving sector is bosonic and carries the $E_6$ gauge quantum numbers, while the left-moving sector is supersymmetric. In the free-field construction both are built from five pairs of free bosons $X^{\pm}_i$ and free fermions bosonized by $H_i$, that is $\Psi^{\pm}_i=e^{\pm iH_i}$, with background charges $a^{\pm}_i$ determined by the BH exponents \cite{Belavin:2025ff1}. In this realization the plus supercurrent of the internal $N=2$ algebra reads
\be
\label{eq:Gplus}
G^{+}=\sqrt2\sum_i e^{iH_i}\big(\partial X^{-}_i-i\,a^{-}_i\,\partial H_i\big),
\ee
and $G^{-}$ is obtained by conjugation. The momenta are quantized on the dual lattices $M(G)$ and $N(G)$, and the fractional parts of the $N(G)$ momenta label the twisted sectors of the orbifold. For a Fermat model the potential is $W_0=\sum_i x_i^{A_{ii}}$, with the diagonal exponents related to the weights by quasi-homogeneity, $A_{ii}\rho_i=d$, and each factor of the internal theory is an $N=2$ minimal model of level $k_i=A_{ii}-2$. For the quintic $A_{ii}=5$ and $k_i=3$. The NS primaries\footnote{We consider the Neveu--Schwarz sector, which carries the bosonic components of the massless multiplets. The Ramond sector supplies their spacetime-fermionic superpartners and is fixed by supersymmetry through the half-unit spectral flow, so it does not require separate counting.} of one factor are labeled by $(l,q)$ with $0\le l\le k$, $q=-l,-l+2,\dots,l$, which gives $\tfrac{(k+1)(k+2)}{2}$ representations, ten for $k=3$, and, writing $A=k+2$,
\be
\label{eq:dimcharge}
\Delta_{l,q}=\frac{l(l+2)-q^2}{4A}\,,\qquad Q_{l,q}=\frac{q}{A}\,.
\ee
The chiral primaries have $q=l$ and the antichiral ones $q=-l$. On the free-field side the basic vertices are the exponentials
\be
\label{eq:vertex}
V_{\vec m,\vec n,\vec S}=\exp\big[\,\vec m\cdot\vec X^{-}+\vec n\cdot\vec X^{+}+i\vec S\cdot\vec H\,\big],
\ee
labeled by the momenta $(\vec m,\vec n,\vec S)$ with $\vec m\in M(G)$ and $\vec n\in N(G)$. The general vertices contain addition polynomial prefactors in $\partial X^{\pm}_i$ and $\partial H_i$. The dictionary between the momenta and the minimal-model labels is $l_i=m_i+A_{ii}n_i$, $q_i=m_i-A_{ii}n_i$, $s_i=-2S_i$ \cite{Makarov:2026ff2}.

A complete massless vertex is a product of a left and a right factor, subject to the mutual locality conditions. Each such pairing of a left chiral primary with a right chiral primary produces one massless $N=1$ chiral multiplet transforming in the $\rep{27}$ representation of $E_6$, and the pairing with a right antichiral primary produces one multiplet in the $\brep{27}$. A singlet is produced by the pairing of a left chiral primary, given by a product of per-factor chiral primaries satisfying
\be
\sum_i \frac{l^L_i}{A_{ii}}=1\,,
\ee
and a right state of total dimension one and total charge zero. The left factor is rigid because it saturates the $N=2$ BPS bound, $\Delta=Q/2$, and is therefore necessarily a chiral primary. All freedom is on the right, and the right states at $(\Delta,Q)=(1,0)$ include descendants. A direct enumeration of the exponential vertices leads to a substantial overcounting. For the quintic it produces over five hundred lattice points with the singlet quantum numbers, while the true number of states is $330$. A lattice point specifies a representation and a grade rather than a state, and the same representation reappears in several lattice frames. The counting must therefore be organized by representations, which is done in the next section.

\section{The universal counting rule from Shapovalov ranks}
\label{sec:modules}

Let $M_{(l,q)}$ be the irreducible NS representation of the level-$k$ minimal model built on the primary \eqref{eq:dimcharge}. The number of its states at level $n$ above the primary with charge $Q_{l,q}+\delta$ equals the rank of the Shapovalov matrix, i.e. the Gram matrix of inner products among descendant states, of the Verma module at the grade $(n,\delta)$. Unitarity guarantees that the rank subtracts every null vector together with all its descendants, including the embedded nulls specific to the minimal model. Indeed, in a unitary theory every null vector is orthogonal to every state of the representation, so the null vectors and their descendants span exactly the kernel of the Gram form at each grade. Passing from the matrix to its rank removes the kernel, and no knowledge of the embedding structure of the null vectors is required. For massless states the total conformal weight of the right vertex is fixed to one. A level-$n$ descendant therefore requires the underlying internal primary to have conformal weight at most $1-n$. Since this weight is non-negative in a unitary theory, only $n\le1$ can contribute.

At level $\tfrac12$ the basis is $G^{\pm}_{-1/2}|l,q\rangle$ and the norms follow from the anticommutators $\{G^{\mp}_{1/2},G^{\pm}_{-1/2}\}=2L_0\mp J_0$ evaluated on the primary. In this formula and in \eqref{eq:gram} below, $\Delta$ and $Q$ denote the primary values $\Delta_{l,q}$ and $Q_{l,q}$ of \eqref{eq:dimcharge} for the factor considered:
\be
\label{eq:norms}
\big|G^{+}_{-1/2}|l,q\rangle\big|^2=2\Delta-Q\,,\qquad
\big|G^{-}_{-1/2}|l,q\rangle\big|^2=2\Delta+Q\,.
\ee
The vanishing norms are exactly the chiral and antichiral null vectors. At level one and charge shift zero the basis is $\{L_{-1},J_{-1},G^{+}_{-1/2}G^{-}_{-1/2}\}$ acting on $|l,q\rangle$ and the Gram matrix is
\be
\label{eq:gram}
\begin{pmatrix}
2\Delta & Q & 2\Delta+Q\\
Q & c/3 & 2\Delta+Q\\
2\Delta+Q & 2\Delta+Q & (2\Delta+Q)(2\Delta-Q+2)
\end{pmatrix},
\ee
with $c=3k/(k+2)$. For $k=3$ the resulting content of the ten representations is collected in table \ref{tab:content}.

\begin{table}
\centering
\begin{tabular}{ccccc}
\hline
$(l,q)$ & $\big(\tfrac12,+1\big)$ & $\big(\tfrac12,-1\big)$ & $(1,0)$ & comment\\
\hline
$(0,0)$ & $0$ & $0$ & $1$ & only $J_{-1}$\\
$(1,1)$, $(2,2)$ & $0$ & $1$ & $2$ & chiral\\
$(3,3)$ & $0$ & $1$ & $1$ & flowed vacuum\\
$(1,-1)$, $(2,-2)$ & $1$ & $0$ & $2$ & antichiral\\
$(3,-3)$ & $1$ & $0$ & $1$ & \\
$(2,0)$ & $1$ & $1$ & $3$ & generic\\
$(3,\pm1)$ & $1$ & $1$ & $2$ & embedded null\\
\hline
\end{tabular}
\caption{Number of states of the irreducible representations of the $k=3$
minimal model at the grades relevant for the massless states. The pairs
$(n,\delta)$ label the level above the primary and the charge shift. Note the entries for
$(3,3)$ and $(3,\pm1)$, which differ from the naive Verma counting.}
\label{tab:content}
\end{table}

Two entries of the table differ from the generic count of three and deserve a comment. The representation $(3,3)$ is the spectral-flow image of the vacuum, so only the image of $J_{-1}|0\rangle$ survives at level one, giving a single state. The representations $(3,\pm1)$ contain an embedded null vector at level one, which reduces the count to two. In both cases the extra null vectors are subtracted automatically by the rank of \eqref{eq:gram}. It is worth stating precisely where these level-one entries act. A level-one descendant carries weight $\Delta+1$, while the total weight of the right vertex is one, so a factor can sit at level one only if its primary has vanishing weight, and the only representation of vanishing weight is the vacuum. The level-one column of the table is therefore reached by the vacuum row alone, and the massless counts of this paper rest on the primaries and on the level-$\tfrac12$ norms \eqref{eq:norms}. This does not make the exceptional representations dispensable: each of the twenty general vertices of section \ref{sec:quintic}, the class that separates the quintic total $330$ from the $326$ of \cite{Belavin:2025ff1}, carries the primary of $(3,3)$ among its five factors, and the primaries of $(3,\pm1)$ enter the twisted sectors of the orbifold spectra of sections \ref{sec:1721} and \ref{sec:new} in the same way. It is only the level-one content of these representations that the massless problem leaves unused, and the rank supplies it at no extra cost for the massive levels.

\section{Twisted sectors and complete vertices}
\label{sec:sectors}
To apply the counting rules derived above, we now construct the orbifold sectors.
An orbifold is obtained by taking the quotient of the tensor product of minimal models by an admissible symmetry group $G$. This requires two ingredients: projecting onto $G$-invariant states and adding the twisted sectors required by modular invariance. The complete spectrum is therefore organized into sectors labeled by
$(\vec l,\vec{\bar t},\vec w)$, where the twist
$\vec w\in G$ specifies the twisted sector. 

Within each sector, the right-moving factor is the untwisted minimal-model primary with
$\bar q_i=l_i-2\bar t_i$,
$0\le\bar t_i\le l_i$.
The left-moving factor is obtained from the same primary by spectral flow with twist $\vec w$, following
\cite{Belavin:2022sf,Belavin:2023bp}. Defining
\[
t^L_i=(\bar t_i+w_i)\bmod A_{ii},
\]
its minimal-model representation is
\be
\label{eq:leftred}
(l^L_i,\,q^L_i)=
\begin{cases}
(l_i,\;l_i-2t^L_i), & t^L_i\le l_i,\\[2pt]
(k_i-l_i,\;k_i-l_i-2t'_i), & t^L_i\ge l_i+1,
\end{cases}
\ee
where the second branch, with $t'_i=t^L_i-l_i-1$, follows from the field identification of the minimal model.

Not every sector contributes physical states.
The complete vertex operator must be mutually local with every generator of $G$, which reduces to
\be
\label{eq:locality}
\sum_i
\frac{g_i(\bar q_i-w_i)}{A_{ii}}
\in\mathbb Z,
\qquad
g\in G.
\ee
Equivalently,
$\vec w^*=\vec{\bar q}-\vec w$
must belong to the dual group $G^*$, reproducing the free-field lattice conditions of
\cite{Makarov:2026ff2}.
The remaining GSO conditions constrain only the total $U(1)$ charges and are automatically satisfied for the vertices considered here.

The massless spectrum consists of three kinds of multiplets, the generations in the $\rep{27}$, the antigenerations in the $\brep{27}$, and the singlets. Each physical field is labeled by the sector it belongs to, so to find the complete list of fields we count the states contributed by each sector and sum over all sectors. How many states a sector contributes depends on the kind of field.

For the $\rep{27}$ and $\brep{27}$ this is immediate, because each sector carries only one such state. Namely, the generations come from the sectors whose right and left factors are both chiral primaries. On the right this requires $\bar t_i=0$, so that $\bar q_i=l_i$, with the tuple at total dimension one half, $\sum_i l_i/A_{ii}=1$. The left factor, after the twist, must likewise be a chiral primary in every factor, with $\sum_i l^L_i/A_{ii}=1$. Every such sector that also satisfies the locality condition~\eqref{eq:locality} produces one $\rep{27}$ multiplet, and the number of generations is the number of these sectors. The antigenerations arise in the same way from the sectors with $\bar t_i=l_i$, whose right factor is an antichiral primary.

For the singlets a single sector can carry several states, because the right factor need not be the primary but may include descendants, as long as the dimensions add up to one and the charges to zero. The number of singlets is therefore
\be
\label{eq:count}
\#\rep{1}
=
\sum_{\text{physical sectors}}
\prod_i
N_{(l_i,\bar q_i)}(n_i,\delta_i),
\ee
where the sum runs over the sectors whose left factor, after the twist, is a chiral primary in every factor, with $\sum_i l^L_i/A_{ii}=1$. For each such sector the right factor can be any of the states whose grades $(n_i,\delta_i)$, added to the primary dimensions and charges, satisfy the following conditions 
\[
\sum_i(\Delta_{l_i,\bar q_i}+n_i)=1,
\qquad
\sum_i(Q_{l_i,\bar q_i}+\delta_i)=0\,.
\]
The number of such states $N_{(l_i,\bar q_i)}(n_i,\delta_i)$ is listed in Table~\ref{tab:content}. In the proposed counting procedure each singlet is taken into account exactly once, unlike a direct enumeration of the exponential vertices, where a lattice point fixes the representation and the grade but not the number of states at that position, which can be zero when a null vector occurs. Taking the twisted left factor to be antichiral instead of chiral, with left charge $Q^L=-1$ instead of $Q^L=+1$, produces the antiparticle spectrum. CPT invariance requires each massless state to be accompanied by an antiparticle of opposite charges, so the two spectra must agree. In our construction the two counts agree in every case, which is a nontrivial check on the sector enumeration and on the content of Table~\ref{tab:content}.

\section{The orbifold with Hodge numbers (17,21)}
\label{sec:1721}

The first test is the quintic orbifold whose spectrum was already computed in \cite{Makarov:2026ff2} by two independent methods. It is one of the exceptional orbifolds of the introduction, where the true Hodge numbers are larger than the deformation counts: here $h^{1,1}=21$ against only $5$ polynomial mirror deformations, the extra $16$ arising from the twisted sectors. Its admissible group has two generators, $G=\langle[1,1,1,1,1],[0,1,1,4,4]\rangle$. Enumerating the local sectors and summing the module content over them, as prescribed in section \ref{sec:sectors}, gives
\be
\#\rep{27}=17\,,\qquad \#\brep{27}=21\,,\qquad \#\rep{1}=234\,,
\ee
in complete agreement with \cite{Makarov:2026ff2}. The $234$ singlets decompose into three classes. The right chiral tuples with a $G^{-}_{-1/2}$ descendant give $53$ states, the number quoted as the first group in \cite{Makarov:2026ff2}. The right antichiral tuples with a $G^{+}_{-1/2}$ descendant give $65$ states. 

The remaining $116$ states are right primaries at $(\Delta,Q)=(1,0)$ in twisted sectors. The examples listed in the appendix of \cite{Makarov:2026ff2} are recognized individually: the one-oscillator example, whose prefactor $\partial X^{-}_2-i\partial H_2$ is produced by the supercurrent \eqref{eq:Gplus} through bosonization, is the $G^{+}$ descendant in the sector $\vec w=(0,1,1,4,4)$, while the two-oscillator examples, despite the extra prefactor, are right primaries rather than higher descendants. The $164$ general vertices of \cite{Makarov:2026ff2} are thereby given a closed description.

\section{The quintic revisited}
\label{sec:quintic}

The quintic is the original example of Gepner construction, which deserves closer examination in the new counting frame, and it is here that our count differs from the value in the literature. Its admissible group is the minimal one, $G=\langle\beta_0\rangle$ with the single generator $\beta_0=[1,1,1,1,1]$. The same procedure gives
\be
\#\rep{27}=101\,,\qquad \#\brep{27}=1\,,\qquad \#\rep{1}=330=305+5+20\,.
\ee
The class of $305$ states consists of the $G^-$ descendants of the right chiral tuples and reproduces the count of \cite{Belavin:2025ff1}, one state for every nonzero exponent of every deformation monomial. The class of $5$ states consists of the $G^+$ descendants of the right antichiral tuple associated with the mirror deformation. The remaining $20$ states belong to the class of general vertices. They are right primaries at $(\Delta,Q)=(1,0)$ in the twisted sector $\vec w=4\beta_0$. In this sector the reduction \eqref{eq:leftred} allows only four right primaries with a chiral left partner, namely $(1,-1)$, $(2,0)$, $(3,1)$ and $(3,3)$, whose chiral left partners carry $l^L=1,2,3,0$ respectively. Imposing $\sum_i\Delta_i=1$ and $\sum_i Q_i=0$ on tuples built from these four leaves a unique solution, three factors $(1,-1)$, one $(2,0)$ and one $(3,3)$, that is the multiset $\{1,1,1,2,3\}$. Its $5!/3!=20$ permutations $\sigma$ label the states, and the right factor is the primary with $l_i=\sigma_i$ and
\be
\bar q_i=\begin{cases}
-1\,, & \sigma_i=1\,,\\
\;\;0\,, & \sigma_i=2\,,\\
\;\;3\,, & \sigma_i=3\,,
\end{cases}
\ee
so that $\sum_i\Delta_i=3\cdot\tfrac{1}{10}+\tfrac25+\tfrac{3}{10}=1$ and $\sum_i Q_i=3\cdot(-\tfrac15)+0+\tfrac35=0$. The lattice data are $\vec w^*=\vec{\bar q}-\vec w$, a permutation of $(0,0,0,1,4)$ modulo $5$, which is the class of the free-field momentum $\vec m^R$. In the free-field language these are pure exponential vertices at twisted momenta. The left partners are the deformation monomials with exponents $1$, $2$, $0$ on the factors with $\sigma_i$ equal to $1$, $2$, $3$.

The total $330$ disagrees with the value $326$ quoted in \cite{Belavin:2025ff1} with reference to \cite{Gepner:1987qi}. Two independent observations show that $326$ cannot be the count at the Gepner point. First, the twenty general vertices form a single orbit of the $S_5$ permutation symmetry of the quintic, and the other classes decompose into orbits of sizes $5$, $10$, $20$ and $30$. No projection compatible with the symmetry can remove four states, so a consistent count keeps $330$ or drops to $310$. Second, four states of exactly this type survive the projection of the $(17,21)$ orbifold and sit inside the $234$ states of the previous section. The rules that give $234$ therefore force $330$.

The resolution is provided by the Landau--Ginzburg computation of Kachru and Witten \cite{Kachru:1993pg}, who counted the massless spectrum of the quintic Landau--Ginzburg orbifold sector by sector and found precisely $330$ singlets,
\be
330=1+101+228\,,
\ee
together with four neutral gauginos completing an enhanced $U(1)^4$, in agreement with the exactly solvable Gepner model at that point. This is exactly our total. Among the $330$ one recognizes the single size (K\"ahler) modulus and the $101$ deformations of the defining polynomial, while the remaining $228$, which include the twenty general vertices, make up the rest. The count $326$ of \cite{Belavin:2025ff1} differs from $330$ by four states. These four are charged under the enhanced $U(1)^4$, which is present only at the Gepner point; away from it the $U(1)^4$ is broken and they acquire masses through the D-terms \cite{Kachru:1993pg,Aspinwall:2011us}. Thus $326$ is the count at a generic point of the moduli space, and $330$ the count at the Gepner point, where the exactly solvable model and the free-field construction are both defined. The total $326$ that \cite{Belavin:2025ff1} identifies with the Gepner result \cite{Gepner:1987qi} is therefore the generic-moduli value, not the count of the Gepner model itself, and the general vertices of the quintic number twenty rather than sixteen.

The quintic total itself is not new: it is the value carried by the Gepner model at its enhanced point, computed by Kachru and Witten. What the free-field construction adds is a uniform counting scheme that, as the next section shows, extends without modification to Berglund--H\"ubsch orbifolds whose spectra have not been computed before.

\section{New orbifold spectra}
\label{sec:new}

With the counting rules validated we compute the spectra of the two remaining quintic orbifolds of \cite{Belavin:2020quintic}. The results are collected in table \ref{tab:spectra}.

\begin{table}
\centering
\begin{tabular}{lccc}
\hline
model & $\#\rep{27}$ & $\#\brep{27}$ & $\#\rep{1}$\\
\hline
quintic & $101$ & $1$ & $330$\\
$Z_5[0,1,1,4,4]$ & $17$ & $21$ & $234$\\
$Z_5[0,1,2,3,4]$ & $21$ & $1$ & $210$\\
$Z_5[0,0,0,1,4]$ & $49$ & $5$ & $258$\\
\hline
\end{tabular}
\caption{The massless spectra of the quintic and its three $Z_5$ orbifolds at
the Gepner point. The singlet counts refer to complex scalars (chiral multiplets).}
\label{tab:spectra}
\end{table}

Two features are worth noting. For $Z_5[0,1,2,3,4]$ the generation count $21$ agrees with the number of invariant deformations, the orbifold is not exceptional in this sector. For $Z_5[0,0,0,1,4]$ the generation count is $49=25+24$. The $25$ polynomial deformations are supplemented by $24$ states arising because the same right chiral tuple appears in several twisted sectors, so the multiplicity over $\vec w$ supplies the states that the untwisted combinatorics misses. This is the free-field realization of the exceptional Hodge number $h^{2,1}=49$, in agreement with the Batyrev--Borisov geometry. It also matches the combinatorial description of \cite{Aleshin:2026abk}, where the same $24$ states appear as the Roan pairs of elements of the mirror groups. The $16$ twisted states of the $(17,21)$ orbifold match the $16$ Roan pairs of \cite{Aleshin:2026abk} in the same way.

All four models, together with their Berglund--H\"ubsch--Krawitz mirrors defined by the dual groups $G^*$ \cite{Berglund:1993fj,Krawitz:2009}, were treated by the same rules. The mirror map of this class is well studied, in particular at the level of the periods of the multiple BHK mirrors \cite{Belavin:2020per}. In every case the mirror spectrum has the $\rep{27}$ and $\brep{27}$ counts interchanged and the same number of singlets, providing an exact and nontrivial check of the construction.

The singlet totals of table \ref{tab:spectra} are Gepner-point counts. As for the quintic, the models possess enhanced $U(1)$ symmetries at the Gepner point, and the parts of the singlet spectra charged under them are expected to acquire D-term masses away from the special point. The geometric interpretation of the new totals deserves a separate analysis.

\section{Conclusions}
\label{sec:concl}

In this work the free-field construction of the heterotic string on Berglund--H\"ubsch orbifolds was completed to the point where it produces the full massless spectrum from the defining data, the potential $W_0$ and the admissible group $G$, alone. The only ingredient beyond the known lattice conditions is the content of the irreducible $N=2$ representations, and it is supplied once and for all by the ranks of the Shapovalov matrices. The general vertices, previously known through examples, are thereby counted by a closed rule and written explicitly.

The analysis also leads to a revised singlet count for the quintic at the Gepner point. The total is $330$, in agreement with the Landau--Ginzburg computation of Kachru and Witten and with the enhanced $U(1)^4$ gauge symmetry, while the frequently quoted $326$ is the geometric count at generic complex structure. The spectra of the two remaining quintic orbifolds are new. Their exceptional Hodge numbers arise in the free-field language as multiplicities over the twisted sectors.

The main significance of the counting rule lies beyond the examples treated here. The rule is model independent. The content tables are computed once for every minimal-model level, the sector structure and the locality condition are fixed by the group, and nothing in the procedure is specific to the quintic. Combined with the combinatorial database of the orbifolds of all $147$ Fermat threefolds considered in \cite{Aleshin:2026abk}, this opens the way to a complete classification of the massless heterotic spectra of Fermat type BH orbifolds, with the Hodge numbers of \cite{Aleshin:2026abk} serving as an independent check of the charged sector and with the singlet spectra computed for the first time. Such a classification would also determine the enhanced gauge symmetries of the special points and the associated Gepner-point singlets, which is exactly the information one needs in order to follow the spectra along the moduli space. For non-Fermat potentials of chain and loop type no minimal-model description exists, and the free-field construction equipped with the present counting rule appears to be the only systematic method available.

Three directions therefore seem natural. The first is to understand how the Gepner-point spectrum evolves along the Calabi--Yau moduli space. In particular, one would like to determine which of the Gepner-point singlets remain massless once the D-term constraints associated with the enhanced $U(1)$ gauge symmetries are imposed. The second is the classification program described above, first for the Fermat class and then for the non-Fermat BH models. The third is the intrinsic proof of modular invariance of the construction, for which the complete and validated spectrum obtained here is a necessary starting point.

\acknowledgments
It is a pleasure to thank Alexander Belavin for useful discussions.

\bibliographystyle{JHEP}
\bibliography{refs}

\providecommand{\href}[2]{#2}\begingroup\raggedright\begin{thebibliography}{10}

\bibitem{Candelas:1985en}
P.~Candelas, G.~T. Horowitz, A.~Strominger and E.~Witten, \emph{Vacuum
  configurations for superstrings}, {\emph{Nucl. Phys. B} {\bf 258} (1985)
  46--74}.

\bibitem{Gepner:1987qi}
D.~Gepner, \emph{Exactly solvable string compactifications on manifolds of
  {SU(N)} holonomy}, {\emph{Phys. Lett. B} {\bf 199} (1987) 380--388}.

\bibitem{Gepner:1988qi}
D.~Gepner, \emph{Space-time supersymmetry in compactified string theory and
  superconformal models}, {\emph{Nucl. Phys. B} {\bf 296} (1988) 757}.

\bibitem{Berglund:1993fj}
P.~Berglund and T.~Hubsch, \emph{A generalized construction of mirror
  manifolds}, {\emph{Nucl. Phys. B} {\bf 393} (1993) 377--391},
  [\href{http://arxiv.org/abs/hep-th/9201014}{{\tt hep-th/9201014}}].

\bibitem{Belavin:2025ff1}
A.~Belavin, \emph{Free field construction of heterotic string compactified on
  {Calabi--Yau} manifolds of {Berglund--Hubsch} type in the {Batyrev--Borisov}
  combinatorial approach}, {\emph{Nucl. Phys. B} {\bf 1018} (2025) 117055},
  [\href{http://arxiv.org/abs/2506.01068}{{\tt 2506.01068}}].

\bibitem{Makarov:2026ff2}
G.~Makarov, D.~Gepner and A.~Belavin, \emph{Free-field construction of
  heterotic string compactified on {Calabi--Yau} orbifolds via correspondence
  with {N=2} {SCFT} minimal models}, {\emph{arXiv} (2026) },
  [\href{http://arxiv.org/abs/2606.27490}{{\tt 2606.27490}}].

\bibitem{Batyrev:1994hm}
V.~V. Batyrev, \emph{Dual polyhedra and mirror symmetry for {Calabi--Yau}
  hypersurfaces in toric varieties}, {\emph{J. Alg. Geom.} {\bf 3} (1994)
  493--535}, [\href{http://arxiv.org/abs/alg-geom/9310003}{{\tt
  alg-geom/9310003}}].

\bibitem{Borisov:2013}
L.~A. Borisov, \emph{{Berglund--Hubsch} mirror symmetry via vertex algebras},
  {\emph{Commun. Math. Phys.} {\bf 320} (2013) 73--99},
  [\href{http://arxiv.org/abs/1007.2633}{{\tt 1007.2633}}].

\bibitem{Belavin:2023bp}
A.~Belavin and S.~Parkhomenko, \emph{{Mirror symmetry and new approach to
  constructing orbifolds of Gepner models}}, {\emph{Nucl. Phys. B} {\bf 998}
  (2024) 116431}, [\href{http://arxiv.org/abs/2311.15403}{{\tt 2311.15403}}].

\bibitem{Aleshin:2026abk}
S.~Aleshin, A.~Belavin and G.~Koshevoy, \emph{{Hodge numbers for orbifolds of
  Calabi-Yau threefolds Ferma type and the Roan pairs}},
  \href{http://arxiv.org/abs/2607.13946}{{\tt 2607.13946}}.

\bibitem{Belavin:2022sf}
A.~Belavin, V.~Belavin and S.~Parkhomenko, \emph{{Explicit construction of N=2
  SCFT orbifold models. Spectral flow and mutual locality}}, {\emph{Nucl. Phys.
  B} {\bf 982} (2022) 115891}, [\href{http://arxiv.org/abs/2206.03472}{{\tt
  2206.03472}}].

\bibitem{Krawitz:2009}
M.~Krawitz, \emph{{FJRW} rings and {Landau--Ginzburg} mirror symmetry},
  {\emph{arXiv} (2009) }, [\href{http://arxiv.org/abs/0906.0796}{{\tt
  0906.0796}}].

\bibitem{Kachru:1993pg}
S.~Kachru and E.~Witten, \emph{{Computing the complete massless spectrum of a
  Landau-Ginzburg orbifold}}, {\emph{Nucl. Phys. B} {\bf 407} (1993) 637--666},
  [\href{http://arxiv.org/abs/hep-th/9307038}{{\tt hep-th/9307038}}].

\bibitem{Aspinwall:2011us}
P.~S. Aspinwall and M.~R. Plesser, \emph{{Elusive worldsheet instantons in
  heterotic string compactifications}}, {\emph{Proc. Symp. Pure Math.} {\bf 85}
  (2012) 33--52}, [\href{http://arxiv.org/abs/1106.2998}{{\tt 1106.2998}}].

\bibitem{Belavin:2020quintic}
A.~Belavin and B.~Eremin, \emph{Mirror pairs of orbifolds of the quintic},
  {\emph{JETP Lett. (Pis'ma v ZhETF)} {\bf 112} (2020) 388--393}.

\bibitem{Belavin:2020per}
A.~Belavin, V.~Belavin and G.~Koshevoy, \emph{{Periods of the multiple
  Berglund-Hubsch-Krawitz mirrors}}, {\emph{Lett. Math. Phys.} {\bf 111} (2021)
  93}, [\href{http://arxiv.org/abs/2012.03320}{{\tt 2012.03320}}].

\end{thebibliography}\endgroup

\end{document}